\documentclass[conference]{IEEEtran}
\IEEEoverridecommandlockouts
\usepackage{cite}
\usepackage{amsmath,amssymb,amsfonts}
\usepackage{algorithmic}
\usepackage{graphicx}
\usepackage{textcomp}
\usepackage{xcolor}
\usepackage{tabularx}
\usepackage{subcaption}
\usepackage{booktabs}
\usepackage{array}

\def\BibTeX{{\rm B\kern-.05em{\sc i\kern-.025em b}\kern-.08em
    T\kern-.1667em\lower.7ex\hbox{E}\kern-.125emX}}

\usepackage{fancyhdr}

\begin{document}

\title{Investigating AI in Peer Support via Multi-Module System-Driven Embodied Conversational Agents}

\author{
    \IEEEauthorblockN{Ruoyu Wen}
    \IEEEauthorblockA{
        \textit{University of Canterbury} \\
        Christchurch, New Zealand \\
        rwe77@uclive.ac.nz
    }
    \and
    \IEEEauthorblockN{Xiaoli Wu}
    \IEEEauthorblockA{
        \textit{University of Canterbury} \\
        Christchurch, New Zealand \\
        xiaoli.wu@pg.canterbury.ac.nz
    }
    \and
    \IEEEauthorblockN{Kunal Gupta}
    \IEEEauthorblockA{
        \textit{The University of Auckland} \\
        Auckland, New Zealand \\
        kunal.gupta@auckland.ac.nz
    }
    \and
    \IEEEauthorblockN{Simon Hoermann}
    \IEEEauthorblockA{
        \textit{University of Canterbury} \\
        Christchurch, New Zealand \\
        simon.hoermann@canterbury.ac.nz
    }
    \and
    \IEEEauthorblockN{Mark Billinghurst}
    \IEEEauthorblockA{
        \textit{University of South Australia} \\
        Adelaide, Australia \\
        mark.billinghurst@auckland.ac.nz
    }
    \and
    \IEEEauthorblockN{Alaeddin Nassani}
    \IEEEauthorblockA{
        \textit{University of Aizu} \\
        Fukushima, Japan \\
        alaeddin@ahlab.org
    }
    \and
    \IEEEauthorblockN{Dwain Allan}
    \IEEEauthorblockA{
        \textit{University of Canterbury} \\
        Christchurch, New Zealand \\
        dwain.allan@canterbury.ac.nz
    }
    \and
    \IEEEauthorblockN{Thammathip Piumsomboon}
    \IEEEauthorblockA{
        \textit{University of Canterbury} \\
        Christchurch, New Zealand \\
        tham.piumsomboon@canterbury.ac.nz
    }
}

\maketitle
\thispagestyle{fancy} 

\begin{abstract}
Young people’s mental well-being is a global concern, with peer support playing a key role in daily emotional regulation. Conversational agents are increasingly viewed as promising tools for delivering accessible, personalised peer support, particularly where professional counselling is limited. However, existing systems often suffer from rigid input formats, scripted responses, and limited emotional sensitivity. The emergence of large language models introduces new possibilities for generating flexible, context-aware, and empathetic responses. To explore how individuals with psychological training perceive such systems in peer support contexts, we developed an LLM-based multi-module system to drive embodied conversational agents informed by Cognitive Behavioral Therapy (CBT). In a user study (N=10), we qualitatively examined participants’ perceptions, focusing on trust, response quality, workflow integration, and design opportunities for future mental well-being support systems.
\end{abstract}

\begin{IEEEkeywords}
Emotion Regulation, Peer Support, Conversational Agent
\end{IEEEkeywords}

\section{Introduction}\label{sec:Introduction}

Large language models (LLMs) offer new possibilities for conversational agents (CAs) used in mental health support. This paper explores how individuals with psychological training perceive LLM-driven CAs designed to deliver methodical and empathetic responses within the context of peer-based mental health support. The global rise in mental health issues presents a major public health challenge. The World Health Organization estimates that one in four people will experience mental health difficulties in their lifetime \cite{world2017depression}. Youth aged 12–24 is a particularly critical period, as most mental disorders emerge during these formative years \cite{patel2007mental}. Early intervention is essential to prevent the escalation of mental health problems and reduce long-term impacts on individuals, families, and communities \cite{rickwood2011promoting,clarke2015systematic}. However, a global shortage of trained mental health professionals continues to limit timely access to care for young people \cite{kakuma2011human,bruckner2011mental}.

Conversational agents (CAs) are increasingly recognised as promising and impactful tools for mental health support \cite{d2020ai}. CAs employ computational linguistics to interpret and respond to user inputs in natural language, engaging users in task-oriented dialogues through text replies or system actions \cite{lester2004conversational}. CAs are widely used in mental health support, with applications such as Woebot, Wysa, and Youper providing conversational therapy and emotional assistance through AI-driven interactions. Traditional CA system rely on rule-based systems, pattern matching, and finite-state machines (FSMs) to manage dialogue \cite{allouch2021conversational}. These systems use predefined rules or patterns to interpret inputs and generate responses, which could only suit narrow domains \cite{adamopoulou2020overview} but limit their ability to handle open-ended or complex conversations.

The rise of LLMs has opened new opportunities for conversational systems, offering advanced natural language understanding and generation capabilities \cite{wu2023brief}. Recent applications such as Dot, Mia, and MoodTalker integrate LLMs to provide more empathetic virtual companionship for mental well-being. At a global scale, the World Health Organization (WHO) launched S.A.R.A.H., an LLM-based online counsellor providing 24/7 health information\footnote{https://www.who.int/campaigns/s-a-r-a-h}
. Yet, the reliability of LLM-generated responses remains a concern: outputs may at times be harmful, inappropriate, or difficult to control \cite{ma2023understanding}. This risk is heightened in sensitive mental health contexts, where LLMs often struggle with nuance and may produce biased or overly generalised responses \cite{ma2024evaluating}.

Through a formative study with experienced psychologists from universities working in the peer support field, we identify the challenges encountered with LLM-generated responses, as well as the decision-making and strategies employed during the process of peer support. Informed by the challenges identified in our formative study and aiming to address the limitations of current LLMs, we developed \textbf{EmoCBT}, a multi-module conversational system inspired by peer-support experts' workflows to drive embodied conversational agents (ECAs), CAs embodied with virtual humans. EmoCBT uses various LLM modules to identify and analyse user inputs, applying different strategies based on the context. Drawing on Cognitive Behavioral Therapy (CBT), a structured and widely adopted approach \cite{fitzpatrick2017delivering,beatty2022evaluating}, the system incorporates emotion regulation techniques to help address users' negative emotions.

Based on EmoCBT, we conducted a user study with ten psychology practitioners role-playing as young people interacting with the ECAs. The study qualitatively examined their perceptions of multi-module AI-driven responses in simulated peer-support scenarios and explored how such systems could be designed and integrated into real-world mental well-being support.

The main contributions of this paper are:

\begin{enumerate}
    \item A formative study identifying challenges and strategies in designing and using LLM-driven systems for peer support in mental well-being.
    \item The design of EmoCBT, a multi-module conversational system integrating CBT to drive ECAs to deliver empathetic, context-aware responses.
    \item A qualitative evaluation with psychology practitioners (N=10) assessing the ECAs, focusing on empathetic peer support interactions.
\end{enumerate}

\section{Related Work}\label{sec:Background}
\subsection {Negative Emotions and Regulation in Young Adulthood}
Young adults navigate a significant life transition, often moving away from established support systems like family while facing new challenges of independent living, work, or study \cite{arnett2000emerging}. This period frequently involves heightened experiences of negative emotions, such as anxiety \cite{kessler2005prevalence}, self-doubt \cite{erikson1968identity, marcia1966development}, loneliness \cite{weiss1973loneliness, laursen2013loneliness}, depression \cite{hankin1998development, brown1978social}, frustration \cite{taylor1991coping}, and a sense of loss of direction  \cite{arnett2000emerging}. Consequently, emotion regulation---the process of managing which emotions one has, when one has them, how they are experienced or expressed \cite{emotionregulation1}---becomes crucial for mental health during this phase.

Common emotion regulation strategies include antecedent-focused approaches like \textit{cognitive reappraisal} (reframing situations to alter emotional impact) and response-focused strategies like \textit{expressive suppression} (inhibiting outward emotional displays) \cite{emotionregulation1}. \textit{Cognitive Behavioral Therapy}, CBT, particularly techniques emphasising cognitive reappraisal, is often used to address cognitive biases linked to negative emotions \cite{young2019positive}. Such CBT principles have been widely integrated into digital interventions, including chatbots and CAs, to offer accessible mental health support \cite{cbt1,cbt2,cbt3,cbt4,cbt5}. Our work draws upon these CBT and emotion regulation principles to inform the design of a CA aimed at supporting users through flexible dialogue.



\subsection{Conversational Agents in Mental Well-being}
CAs are increasingly utilised in mental well-being contexts. Research commonly explores their application in four key areas:
\begin{enumerate}
    \item \textbf{Assessment:} Administering self-report measures (e.g., PHQ-9) to assess mental health symptoms, potentially substituting for face-to-face assessments \cite{sa1}.
    \item \textbf{Intervention:} Delivering therapeutic content, such as CBT, counseling, social skills training, or self-management support \cite{int1,int2,int3,int4,meng2023mediated}, often functioning as tutors, coaches, or healthcare partners \cite{ECAscoping}.
    \item \textbf{Companionship:} Providing long-term emotional support and interaction through agents like Replika \cite{com2}, Mitsuku \cite{com3}, and Xiaoice \cite{xiaoice,loveys2020effect}.
    \item \textbf{Training:} Simulating patient or peer interactions to help develop clinical or peer-support skills \cite{tra1,tra2,tra3}.
\end{enumerate}

A significant focus within CA research is enhancing their ability to perceive and respond appropriately to user emotions. \textit{Sentiment analysis} is often employed to interpret emotional tone from text, with studies indicating that CAs demonstrating emotional interpretation are perceived as more emotionally intelligent \cite{content3}. Beyond text, \textit{acoustic features} from speech have been integrated; for instance, speech emotion recognition (SER) systems have shown potential in reducing negative affect \cite{sound1}. \textit{Multimodal systems} further combine inputs like facial expressions, voice stress, and semantic content to achieve a more nuanced emotional understanding \cite{multi1, multi2}. Frameworks like Yalçın’s \cite{multi3}, which categorises empathy into levels (e.g., mimicry, affect regulation, context-aware reasoning), provide structured approaches for designing more empathetic agents.

\textbf{Embodiment}, representing CAs as virtual humans or characters, is another crucial design dimension, particularly for fostering emotional connection \cite{cassell2001embodied,bickmore2005social,louwerse2009embodied}. Embodied conversational agents (ECAs) can leverage non-verbal cues (facial expressions, gaze, gestures) alongside verbal responses to convey empathy and attentiveness more effectively than text- or voice-only systems \cite{ter2020design}. Research suggests these multimodal cues enhance user trust, perceived warmth, and conversational naturalness, especially in sensitive contexts \cite{cassell2001embodied, yalccin2020empathy,yalccin2018modeling}. Informed by this background, our research explores how LLM-driven analysis of user emotion \cite{wen2024large, elyoseph2023chatgpt} can guide response generation within an embodied agent, aiming to enhance empathetic interaction.

\subsection{Large Language Model Empower Mental Well-being Conversational Agents}
Recent advances in LLMs have significantly enhanced the capabilities of CAs, enabling more natural, responsive, and context-aware interactions for mental well-being support. For instance, a study involving 622 young people found that while AI-generated responses were preferred for general topics, human responses were still deemed more appropriate for highly sensitive issues like suicidal ideation, highlighting nuances in user acceptance \cite{young2024role}.

Several systems demonstrate the integration of LLMs in mental health applications. ComPeer uses LLMs for proactive peer support, focusing on conversational planning, emotion detection, memory reflection, and empathetic response generation, with users reporting increased engagement and reduced stress\cite{liu2024compeer}. Psy-LLM assists mental health professionals by generating answers to psychological queries using LLMs fine-tuned on relevant datasets, showing coherence but facing difficulties with complex topics \cite{lai2023psy}.

Despite progress, significant challenges remain. LLMs can still produce inappropriate, harmful, or overly generic responses, particularly concerning sensitive or nuanced topics like LGBTQ+ issues \cite{ma2024evaluating}. Ensuring long-term memory, consistent persona, and user trust over extended interactions also remains problematic \cite{croes2021can, ma2019exploring}.

Addressing these limitations requires careful system design. Our work introduces \textbf{EmoCBT}, a multi-module system architecture designed to drive ECAs for mental well-being support. It aims to generate contextually appropriate and empathetic responses by integrating CBT principles and leveraging multiple LLM components, specifically tackling some of the identified challenges related to response quality and sensitivity in peer-support contexts. The subsequent sections detail the design of EmoCBT and its evaluation with psychology practitioners.

\section{Formative Study}\label{sec:proto}
We conducted a formative study with four psychologists (three female, one male; mean age = 30.3, SD = 4.65) working in university student support teams. All had at least one year of peer-support experience and formal training.

\begin{figure}[h]
\centering
    \includegraphics[width=0.9\linewidth]{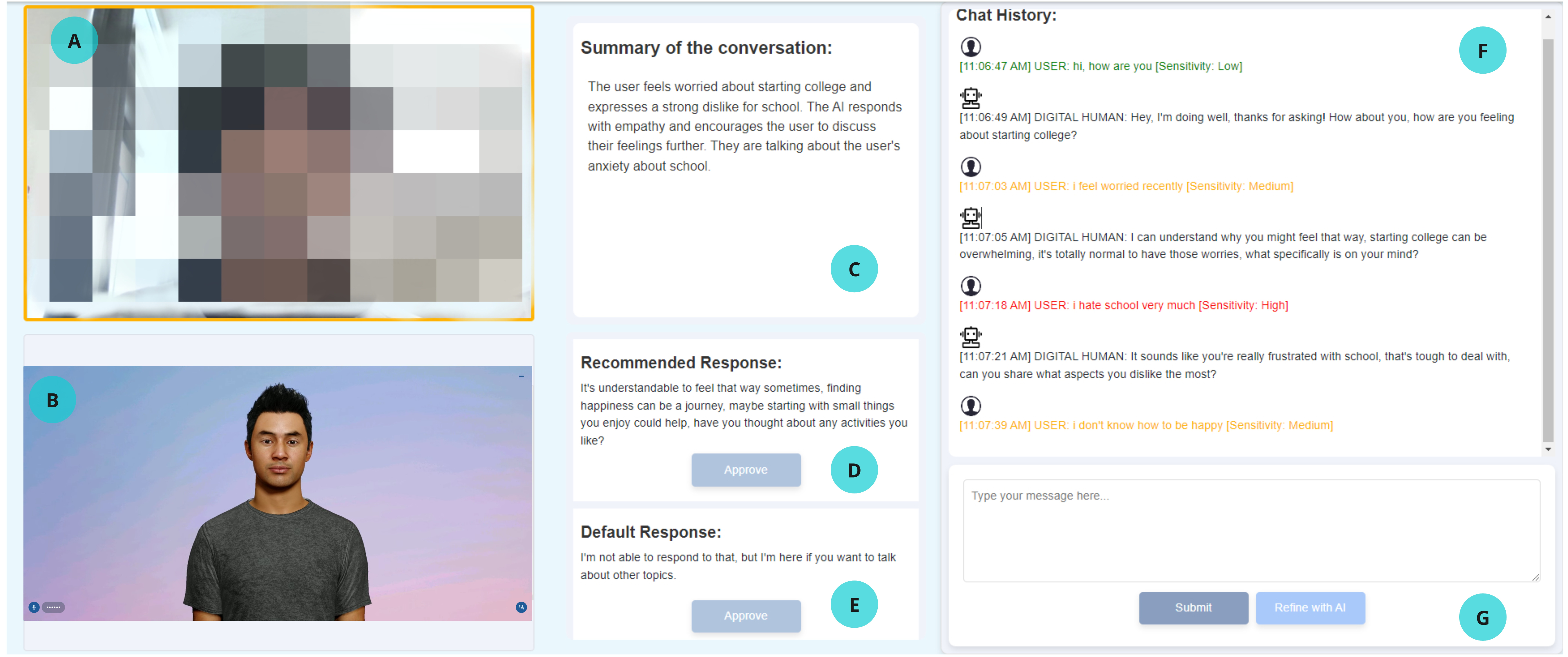}
    \caption{User Interface of the dialogue system: (A) Real-time monitoring screen of the human participant (blurred for anonymity), (B) Real-time monitoring screen of the Virtual Human, (C) Summary of the current conversation, (D) Recommended response, (E) Default response, (F) Chat history of the current conversation, (G) User input interface.}
    \label{fig:ui}
\end{figure}

The study took place in a controlled laboratory setting. Psychologists acted as human-in-the-loop moderators using an AI-supported dialogue system (Fig. \ref{fig:ui}) to mediate conversations between a virtual human and an actor role-playing a stressed first-year student. They reviewed outputs from two LLM-based modules: a detection module that classified input as green (non-stressful), yellow (moderately stressful), or red (highly stressful), and a response module that generated suggestions based on emotional state and dialogue history. Psychologists could approve, edit, or replace these responses before delivery. The system was designed, following prior work \cite{abbas2020crowd}, to maintain human oversight in mental well-being interactions with conversational agents.

In this setup, psychologists moderated the interaction, evaluating both the system’s stress detection and the appropriateness of AI-generated responses to ensure that the conversational agent (CA) remained emotionally sensitive, safe, and consistent with peer-support norms. Their oversight helped preserve the CA’s supportive role and reduced risks from generic or emotionally detached outputs. This process also enabled psychologists to assess the accuracy and effectiveness of LLM-generated content.

After each session, we conducted semi-structured interviews to gather feedback on the multi-module system, challenges encountered, and peer-support workflows. Psychologists reported high satisfaction with the detection module, particularly in handling low-stress input, which usually matched their own judgments. For non- and moderately stressful messages, AI responses were often approved without change. However, high-stress content was frequently edited or rewritten, as participants found the original outputs vague, lacking empathy, or overly “robotic.”

Participants also reflected on their own support practices, which emphasised attentive listening, emotional assessment, boundary clarification, and structured reflection, often drawing on CBT principles. They highlighted the importance of keeping the system within peer-support boundaries, avoiding diagnostic or clinical advice, and ensuring clear referral pathways to professional services when needed.

Based on these findings, we derived three design goals to guide the development of EmoCBT, a multi-module CA system for peer-based mental well-being conversations:

\begin{itemize} 
\item \textbf{DG1: Enhance Detection of Negative Emotions.} Improve the system’s ability to accurately recognise and classify users' emotional states, providing actionable insights that support emotional awareness and regulation.

\item \textbf{DG2: Provide Thoughtful and Structured Support.} Incorporate CBT-informed techniques to generate responses that are empathetic, reflective, and analytically grounded, particularly for emotionally charged situations.

\item \textbf{DG3: Define System Scope and Offer Resources for Severe Cases.} Clearly communicate the system’s boundaries and ensure users are directed to professional mental health resources when their needs exceed what peer support can provide (Highly stressful situations). 
\end{itemize}

\section{System Design}\label{sec:from}
Following the design goals and inspired by previous research \cite{park2023generative,liu2024compeer}, the system involves multiple LLM-driven modules, all using GPT-4o. The core modules are designed to detect emotional cues in real time, adapt responses to emotional states, and guide users through structured CBT-based interactions when needed. The system automatically manages non-stressful interactions and addresses emotional distress when detected, categorising it as \textit{Yellow} or \textit{Red} to ensure appropriate intervention. The following subsections detail each module’s specific function in supporting effective, responsive conversation.

\subsection{Detection Module}
The detection module refines the sensitivity detection approach from the formative study, using clearer colour definitions provided by an external mental counsellor.  Distinct strategies are applied based on the sensitivity category of user messages:

\begin{itemize}
    \item \textbf{Green}: Indicates non-stressful input, allowing the system to continue with casual conversation.
    \item \textbf{Yellow}: Suggests potential emotional distress, prompting further inquiry from the system.
    \item \textbf{Red}: Flags serious psychological distress, leading the system to provide external resources for support.
\end{itemize}

The categorisation was developed based on a guideline created by an experienced mental health counsellor working with young adults.  This guideline was transformed into a system prompt used by the LLM to assess message risk.  While the mediator had the option to override these classifications, no such edits were made during this study.  Mediators were primarily responsible for approving or editing the system’s proposed responses before they were delivered by the conversational agent (CA), thus maintaining oversight and safety in sensitive situations.
\subsection{Empathetic Response Module}
For non-stressful input (\textit{Green}), the system generates responses from a friendly persona characterised as a ``supportive friend, always positive and empathetic.'' The aim is to gently encourage users to share their feelings without appearing intrusive. The \texttt{generate\_empathetic\_response} function utilises conversation history to produce contextually appropriate replies that maintain an open and supportive tone. 

\subsection{Emotion Recognition Module}
For inputs categorised as \textit{Yellow}, the system initiates a deeper emotional analysis through the \texttt{detect\_emotions} function. This module prompts the LLM to extract emotional expressions from user input, focusing on negative emotions commonly experienced by young people, such as anxiety \cite{arnett2000emerging, kessler2005prevalence}, self-doubt \cite{erikson1968identity, marcia1966development}, loneliness \cite{weiss1973loneliness, laursen2013loneliness}, depression \cite{hankin1998development, brown1978social}, frustration \cite{taylor1991coping}, and a sense of loss of direction \cite{arnett2000emerging}. If the user input is brief, the system first prompts for more details to prevent misinterpretation before conducting a full emotion detection. This capability enhances the detection of negative emotions, fulfilling \textbf{DG1}.


\subsection{Deep Conversation Module}
Guided by \textbf{DG2}, we developed a Deep Conversation module. Once the user identifies a specific emotion, the system transitions to a CBT-based dialogue to encourage emotional reflection and exploration. The AI agent dynamically adapts to the user’s needs by selecting appropriate CBT techniques, ensuring that interactions feel personalised. We adopted CBT due to its well-established protocol and its extensive use in prior HCI research involving mental health support \cite{he2023conversational}.


\subsection{Crisis Intervention and Resource Provision}
To fulfil \textbf{DG3}, we introduced a \textbf{Crisis Intervention and Resource Provision} mechanism. For inputs categorised as \textit{Red}, indicating severe emotional distress, the system ceases the dialogue and instead provides external resources, such as helplines and counselling services, directing the user towards appropriate professional help.


\subsection{Memory Module}
A memory system is implemented using conversation history, where conversation logs are stored in a message list for context retrieval. This enables the system to access the full conversation history during interactions, generating more coherent and personalised responses based on previous exchanges. By maintaining context continuity, the system enhances the overall conversational experience.

\subsection{An Example User Case}
Consider Sarah, a first-year university student feeling overwhelmed by upcoming exams and presentations. She begins a conversation with the EmoCBT-driven CA, saying, ``I’m feeling really nervous about my upcoming exams and presentations. I don’t know if I can handle all of this.'' The system processes her input using the detection module and classifies her emotional tone as \textit{Yellow}, indicating moderate stress related to anxiety and nervousness.

The CA responds empathetically: ``It’s completely normal to feel nervous about exams and presentations. University can be stressful, especially in your first year, but I’m here to help. Can you tell me more about what’s been on your mind?'' Encouraged, Sarah shares, ``I feel like I’m going to mess everything up. Every time I think about speaking in front of people, I freeze up.''

Using its emotion recognition module, the system identifies anxiety and self-doubt as prominent emotions and reflects this back: ``From your message, it sounds like you’re feeling both anxiety and self-doubt. Which would you like to focus on first?'' When Sarah chooses anxiety, the CA shifts into a CBT-informed dialogue, prompting reflection with: ``What do you think makes public speaking feel so anxious for you?''

Sarah replies, ``I’m afraid I’ll mess up, and everyone will judge me.'' Instead of offering prescriptive advice, the system supports cognitive reframing: ``It sounds like you’re focusing on what might go wrong. What if you thought about how preparation could reduce some of that pressure?''

Throughout the conversation, the CA adapts its responses based on context, maintaining a conversational tone and encouraging open reflection. It avoids rigid scripts while using CBT techniques to guide the dialogue. The system tracks conversation history, enabling continuity and reinforcing key patterns in Sarah’s thinking. By the end of the session, Sarah reports feeling more confident about managing her anxiety and is encouraged to return if she needs further support.

\begin{figure*}[h]
    \centering
    \includegraphics[width=0.7\linewidth]{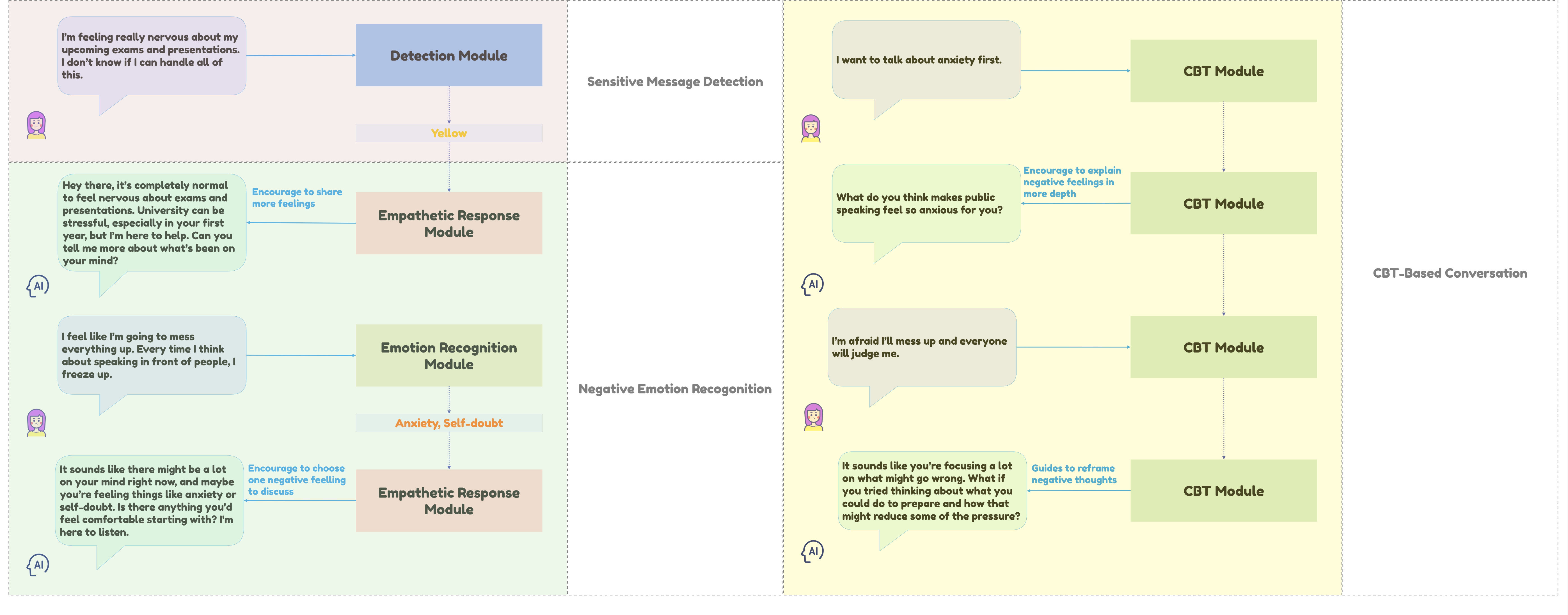}
    \vspace{-0.3cm}
    \caption{User Case: an example of a stressed first-year student talking with EmoCBT about her negative feelings in school. The figure shows the workflow of EmoCBT to support the user's emotional regulation.}
    \label{fig:usercase}
\end{figure*}


\section{User Study}\label{sec:User}
Building on the EmoCBT system, we conducted a within-subjects qualitative study to explore psychology practitioners’ perceptions of LLM-driven conversational agents in peer-support contexts.

\subsection{Research Questions}\label{RQ}
The study was guided by the following research questions:

\begin{enumerate}
    \item[\textbf{[RQ1]}] How do psychology practitioners perceive the use of LLM-driven conversational agents in peer-support conversations?
    
    \item[\textbf{[RQ2]}] How do different design configurations of conversational agents---such as embodiment and adherence to structured protocols---affect practitioners’ perceptions of the peer-support interaction experience? 
    
    \item[\textbf{[RQ3]}] How do psychology practitioners envision the integration of systems like EmoCBT into their future workflows for supporting young people’s mental well-being?
\end{enumerate}

\subsection{Participants} 
Psychological practitioners were recruited to enable safe evaluation of CAs in peer-support contexts. With peer-support training and experience addressing issues faced by young people, they were well-positioned to assess the system’s responses and limitations. Ten practitioners from a local university psychology department were recruited through snowball sampling, social media outreach, and campus posters. Eligibility required prior experience in mental health research, social work, peer support, or related training.  

The sample included six females and four males. Half held bachelor’s degrees and half held master’s degrees. Participants were distributed across three age groups: 18--25 years ($n=4$), 26--35 years ($n=3$), and 36--45 years ($n=3$). Five reported more than three years of mental health experience, two had one to three years, and three had less than one year. Forty percent had previously used mental health applications or chatbots. Familiarity with CBT was high, with 40\% describing themselves as \textit{very familiar}, 50\% as \textit{somewhat familiar}, and 10\% as \textit{not familiar}. In addition, 80\% reported prior experience with emotion regulation techniques. Each participant received \$10 compensation.

\subsection{Study Procedure}

Before the study, participants received an overview of the study, signed a consent form, and were trained to familiarise themselves with the system interface and interaction process. Participants were then asked to role-play a first-year student experiencing emotional distress, while the conversational agent acted as a peer providing emotional support.

The study qualitatively investigated two design features: \textit{Model Type} (Generic vs. EmoCBT) and \textit{Embodiment} (Embodied vs. Voice-only). Each participant experienced all four conditions, with the order of conditions fully counterbalanced across participants to mitigate order effects. Each interaction lasted approximately 5-7 minutes.

\begin{itemize}
    \item \textbf{Generic + Embodied (G1)}: A baseline LLM-driven conversational agent presented via a Soul Machines virtual human~\footnote{https://www.soulmachines.com/}.
    \item \textbf{EmoCBT + Embodied (E1)}: A CBT-informed conversational agent presented via a Soul Machines virtual human.
    \item \textbf{Generic + Voice-only (G0)}: The same baseline agent presented via ChatGPT’s Advanced Voice Mode, without embodiment.
    \item \textbf{EmoCBT + Voice-only (E0)}: The CBT-informed agent presented via voice-only interaction.
\end{itemize}

After each condition, participants completed a short questionnaire and participated in a brief semi-structured interview to reflect on the interaction. They were asked about the CA's empathy, conversational structure, emotional appropriateness, and overall support quality during that condition. The entire session concluded with a final, in-depth semi-structured interview to gather comparative feedback across all conditions.

\subsection{Measures}

Quantitative ratings were collected after each condition, where participants rated their experience on a 0–10 scale across three dimensions: trust in the conversational agent (0 = no trust at all, 10 = complete trust), perceived quality of the conversation (0 = very poor, 10 = excellent), and how well the conversation addressed their concerns (0 = not at all, 10 = completely).

Qualitative data were gathered through the final semi-structured interviews, conducted after participants had experienced all conditions.  These interviews aimed to gain deeper insights into participants’ overall perceptions. Topics included: overall experience with CAs in mental health contexts; preferences between the Generic and EmoCBT models (including aspects like emotion recognition and structured support); perceived effectiveness in problem-solving; trust in the CA’s responses; and the impact of virtual human embodiment. Participants also reflected on any insights or emotional clarity gained from the interactions, and how such systems might integrate into real-world mental health workflows.

Two researchers independently coded the interview transcripts and resolved discrepancies through discussion. A shared codebook was developed and refined iteratively to ensure that resulting themes accurately captured participant perspectives.


\section{Results}\label{sec:Results}

This section presents the study findings, starting with descriptive statistics from participant ratings (trust, quality, effectiveness) across model type and embodiment. We then detail qualitative interview findings on perceptions of LLM agents in peer support, concluding with suggestions for future system integration. 

\subsection{Descriptive Statistics}

We analysed participants' trust ratings (0–10) across two main design features: 1) Model Type (Generic model [G] vs EmoCBT [E]), and 2) Embodiment (Voice-only [0] vs Virtual Human [1]). This resulted in four conditions: G0, G1, E0, and E1. 


\textbf{Model Effect.} Comparing EmoCBT (E0, E1) versus Generic (G0, G1) conditions, EmoCBT showed a slight advantage. Trust averaged 7.45 ($SD=1.50$) for EmoCBT versus 7.1 ($SD=1.77$) for Generic. Perceived quality was slightly higher for EmoCBT ($\bar{x}=7.3$) than Generic ($\bar{x}=7.2$). Effectiveness was rated higher for EmoCBT ($\bar{x}=7.8, SD=1.32$) compared to Generic ($\bar{x}=7.55, SD=1.79$).

\textbf{Embodiment Effect.} Comparing embodied (G1, E1) versus voice-only (G0, E0) condition, voice-only interactions received slightly higher average ratings. Trust averaged 7.5 ($SD=1.76$) for voice-only versus 7.05 ($SD=1.50$) for embodied. Perceived quality was also higher without embodiment ($\bar{x}=7.5, SD=1.54$) than with it ($\bar{x}=7.0, SD=1.81$). Effectiveness ratings followed this trend ($\bar{x}=8.15, SD=1.39$ voice-only vs. $\bar{x}=7.2, SD=1.61$ embodied).

\subsection{In-depth Interview}
Post-study interviews examined practitioner perceptions of the CA’s performance in peer-support contexts across conditions.

\subsubsection{Perceived Benefits and Opportunities}
Participants expressed cautious optimism about LLM-driven CAs, noting benefits such as availability, consistency, and a non-judgmental tone for users with mild emotional concerns. Several observed that the system could help individuals articulate emotions, reflect, and feel heard, especially when structured around CBT principles. Freeform supportive responses and reflective prompts were preferred over direct advice, as they better matched peer-support norms. As P6 described, the interaction felt like “\textit{a structured conversation that helped me slow down and think more deeply.}”

\subsubsection{Challenges and Concerns}
Concerns centered on emotional sensitivity, rigidity, and trust. Some found the CA’s style overly formal or repetitive, which felt impersonal or intrusive (P3: “\textit{It kept asking about my emotions even after I already answered.}”). Others stressed that users in distress may prefer warmth and companionship over structured guidance (P2: “\textit{When someone is struggling, they may not want instructions from a virtual friend—they just want to feel someone’s there.}”).

\subsubsection{Integration into Peer-Support Workflows}
Psychologists viewed LLM-driven agents as complementary rather than replacements for human supporters. They envisioned roles in initiating reflection, extending support outside working hours, or easing transitions to human interaction (P8: “\textit{This could be a helpful first layer—it helps people open up before talking to someone.}”). Some also noted potential as a training tool for peer supporters. Integration suggestions included more adaptive dialogue, clearer topic boundaries, and customisation of the agent’s appearance and voice.

\subsubsection{Protocol Structure: Openness vs. Guidance}
Participants distinguished between models. The Generic CA was seen as relaxed and user-led (P4: “\textit{It helps users open up—it feels less formal.}”), while EmoCBT was described as more structured and guiding (P6: “\textit{It helped me expand on my thoughts without feeling lost.}”). However, some found EmoCBT’s probing overwhelming or fatiguing (P7), or felt it overemphasised negative aspects (P9).

\subsubsection{Embodiment: Trust, Empathy, and Comfort}
Views on embodiment were mixed. Some valued non-verbal cues that enhanced empathy and presence (P7: “\textit{a friendly face that makes it easier to talk}”), while others found them distracting or mismatched to their identity (P10) or even distressing when negative expressions appeared (P9). Cultural fit also influenced comfort (P6). Voice-only interaction was often described as less pressuring and more neutral, with some participants noting it allowed them to “\textit{imagine someone they trusted}” (P2). This modality was seen as particularly suited for sensitive conversations.

\section{Discussion}\label{sec:Discussion}
\subsection{Prompting Mental Well-being: Designing Conversational Agents for Peer Support}
Practitioner feedback indicates that LLMs show promising potential for CAs used in peer support, particularly for mild mental well-being concerns. Traditional CBT-based CAs \cite{cbt1} often struggle to adapt to users’ evolving needs, especially during prolonged or repeated interactions. In contrast, multi-module LLM CAs like EmoCBT can detect emotional nuances in user input, maintain context across interactions, and generate more personalised and empathetic responses. Additionally, LLMs facilitate more natural, open-ended dialogue, reducing friction caused by predefined input constraints \cite{cbt2, cbt4}. The conversational agent's workflow and persona can also be easily customised through prompt engineering, making the system more accessible and user-friendly, especially for psychologists who may lack programming expertise.

Practitioners also envisioned embedding systems like EmoCBT into routine mental-health practice. In interviews, participants highlighted that, when live support is unavailable, the agent could deliver 24/7 peer-like assistance to address mild emotional distress. They further recommended using EmoCBT as a digital journaling platform, enabling therapists to review interaction logs and track clients’ day-to-day emotional trajectories. Future work will examine integration strategies within clinical workflows and evaluate the system’s feasibility and impact through real-world deployment studies

\subsubsection{Structured Protocol}
CAs following a structured protocol, such as EmoCBT, were seen as helpful in guiding users to reflect on their emotions more deeply, offering a clear and supportive conversational flow. However, consistent with prior research \cite{cbt3}, such agents were also sometimes perceived as less empathetic. Their fast-paced, question-heavy responses could feel overwhelming or robotic, reducing user trust \cite{wester2024chatbot}. In contrast, free-form generic agents enabled more user-led interactions, offering support without steering the dialogue. A potential improvement could involve a decision module that dynamically adjusts conversational styles based on the user’s emotional state and conversational context. Additionally, while current LLMs rely solely on verbal input, professional mental health assessments often rely on non-verbal cues, such as facial expressions, posture, tone of voice, and body language, which can reveal hidden emotional states that may contradict spoken words \cite{hsee1992assessments}. Integrating multimodal AI modules that analyse both verbal and non-verbal cues could improve EmoCBT's ability to interpret users' emotional states more accurately and respond with greater empathy.

\subsubsection{Embodiment}
Prior work suggests that embodied CAs, such as virtual humans, can enhance engagement and trust in mental well-being support by combining verbal and non-verbal behaviours \cite{ter2020design}. In our study, younger participants echoed this finding, reporting that facial expressions and listening gestures helped build trust and improved their experience. However, older participants expressed discomfort, citing feelings of pressure or stigma when interacting with embodied agents. This may relate to generational differences in technology familiarity \cite{hamelmann2018impacts, hauk2018ready}, with younger users—Digital Natives—being more accustomed to digital interfaces. In contrast, older users may rely more on nuanced emotional cues and place greater value on relational depth \cite{thomas2007development, carstensen1992social}, making subtle mismatches in virtual expressions more disruptive. These findings highlight the need for age-sensitive personalisation in designing embodied CAs for peer support.

\subsection{Ethical Considerations}
While CAs are accepted for some mental health concerns, users generally prefer human professionals for more complex needs~\cite{cbt5}. In our study, some participants criticised EmoCBT for overemphasising negative emotions, raising concerns about accountability for potential harm. To address this, we propose a human-in-the-loop model: with user consent, professionals oversee responses, allowing autonomous replies for low-stress input, moderator review for moderate-stress, and direct intervention for high-stress cases~\cite{lee2020designing}.

We emphasise that human contact remains essential and do not intend for AI to replace professional support. Rather, the system is designed as a supplementary resource when immediate human assistance is unavailable, with professional oversight ensuring accountability in sensitive cases. In real-world NGO settings, such an approach could enable a single practitioner to supervise multiple conversations and intervene only when escalation is required, providing a scalable balance between access and safety.

Despite these safeguards, privacy concerns persist, as conversations with mental health CAs often involve sensitive personal information~\cite{mireshghallah2024trust}. To further support user safety and trust, we consider the integration of a locally deployed LLM-based anonymiser, which can remove sensitive data before transmitting input to cloud-based LLM services~\cite{wiest2024anonymizing}.

\section{Limitations and Future Work}
This study has several limitations. The agent’s appearance, voice, and persona were predefined, which may have constrained participants’ engagement and trust. The sample size was small and limited to practitioners with peer-support training, reducing the generalisability of findings. While age and cultural background appeared to influence perceptions, these factors were not systematically examined. The short interaction duration also prevented assessment of longitudinal trust-building and support dynamics, which usually emerge over extended use. Although no LLM hallucinations were observed in this study~\cite{ye2023cognitive}, the risk of such issues may grow with longer-term interactions.

Future work will support user customisation of the agent’s appearance, voice, and persona, and move from closed-source APIs to locally deployed models to improve privacy and system control. We plan to recruit a larger, more diverse sample of mental health professionals through online studies and remote deployments, and to conduct longitudinal studies with real users to examine sustained trust, emotional regulation, and risks such as hallucinations. To further address safety concerns, we will implement validation mechanisms and maintain human oversight for high-risk interactions in mental health contexts.

\section{Conclusion}\label{sec:Conclusion}
This study investigates how individuals with psychological training perceive the use of LLM-driven conversational agents in peer support. We developed EmoCBT, a multi-module system for emotion regulation, and conducted evaluations with ten practitioners. Findings reveal perceived benefits, concerns around empathy and trust, and highlight the need for personalisation and human oversight in future design.

\newpage
\section{Ethical Impact Statement}
This study involved participants who had completed peer support training and were asked to simulate stressed first-year university students during their interactions with a conversational agent. The goal was to evaluate the quality and appropriateness of AI-generated peer support responses in a controlled setting. Ethical approval for this study was obtained from the university's ethics committee ([anonymized] Institutional Review Board), and all participants provided informed consent before participating. Each participant received a \$10 voucher, aligned with institutional guidelines for participant compensation.

Potential risks, such as the AI providing misleading or emotionally inappropriate responses, were identified and evaluated through structured feedback from trained peer supporters. The system is not designed for clinical use and is intended to assist with everyday peer-level support; however, its potential misuse in emotionally sensitive scenarios was considered and monitored.

Limitations of the study include the small sample size and the use of a single language context (English), which may affect generalizability. Cultural norms, language proficiency, and individual communication styles may influence how AI responses are perceived, and future work will explore these factors in more diverse settings.


\newpage
\bibliographystyle{IEEEtran}
\bibliography{references}

\end{document}